\documentstyle[aps,prl,epsfig]{revtex}
\tightenlines
\begin{document}
\twocolumn[\hsize\textwidth\columnwidth\hsize\csname@twocolumnfalse\endcsname
\draft
\title{Protien folding simulations with Interacting Growth Walk model}
\author{S.L. Narasimhan$^{*}$,  P.S.R. Krishna, M. Ramanadham}
\address{Solid State Physics Division,  Bhabha Atomic Research Centre,
\\ Mumbai - 400 085, India}
\author{K. P. N. Murthy}
\address{Materials  Science  Division, Indira Gandhi Centre for Atomic Research,
\\ Kalpakkam - 603 102, Tamilnadu, India}
\author{R. Chidambaram}
\address{Central Complex, Bhabha Atomic Research Centre,
\\ Mumbai - 400 085, India}
\maketitle

\begin{abstract}

We demostrate that the recently proposed interacting growth walk (IGW) model, modified for 
generating self-avoiding heteropolymers, proves to be a simpler alternative to the other Monte Carlo 
methods available in the literature for obtaining minimum energy conformations of lattice proteins. In fact,
this simple growth algorithm seems to be capable of quickly leading to low energy states for all the 
three dimensional bench mark HP-sequences investigated.
 
\end{abstract}
\pacs{87.15.By, 02.70.Lq, 87.10.+e}
\vfill
\twocolumn
\vskip.5pc]
\narrowtext
Proteins are non-branching heteropolymers obtained from twenty commonly occurring 
amino acids. Their specific biological functions are intimately related to their 'native'
({\it i.e.,} unique and thermodynamically stable) conformational structure. One of the
most challenging problems in Biophysics is to understand the kinetic mechanism by which the
information encoded in an amino acid sequence helps the protein grow or fold quickly
into its native conformation [1]. Since the amino acids can be classified broadly into
hydrophobic (H) or polar (P) types, the problem may be simplified by treating a protein 
as a random copolymer made up only two types of amino acids with non-bonded
contact interactions between them. If we ignore the intra-molecular structures of the
individual amino acids ({\it i.e.,} treat them as monomer units) then these interactions
can be expressed in terms of a few effective parameters. Such a coarse-grained approach
to this problem is justified by the observation that the protein-folding mechanisms and
rates are determined more by the topology of protein conformations than by the details
of interatomic interactions [2]. By assuming further that the individual bonds in the linear
chain of monomers are all of equal length and can only be along the directions of a 
regular lattice, we have the lattice protein models [3], called HP-models, which are
amenable to exact enumeration as well as Monte Carlo studies.

In this paper, we consider the problem of finding the native conformational structure that
corresponds to a given HP-sequence and contact interactions. For short chains, the
native state (or equivalently, the ground state) can easily be found by an exact enumeration
of all possible conformations. However, for long chains, specially designed Monte Carlo
methods [4] have to be used for generating a large ensemble of conformations before
a ground state search is performed. This ignores the phenomenology of chain growth 
and its conformational relaxation.

The information for building the proteins is transcribed from the DNA to the {\it m}RNA,
and translated into the corresponding amino acid sequences by a specific biochemical
process - namely, as the ribosomes move along the {\it m}RNA strand from one codon 
to the next, different {\it t}RNA molecules lock into the appropriate places one after another, 
resulting in the transfer of the corresponding amino acids to the growing protein molecule. 
And as the protein grows, it coils into its secondary and tertiary structures as well.

In this paper, we discuss an implementation of this growth scenario for any given HP-sequence
using the recently proposed Interacting Growth Walk (IGW) model [5]. We demonstrate
that this simple growth algorithm is capable of quickly finding the ground state conformations 
for all the three dimensional bench mark sequences investigated [6,7].

Given a sequence of H and P monomers, say ${\cal S}_{N} = \{ A_{j} = H$ or $P; j = 0,1, ..., N-1\}$,
we start growing a lattice protein by placing the first monomer $A_{0}$ at an arbitrarily
chosen site, ${\bf r}_{0}$, of a regular $d$-dimensional lattice of coordination number $z$.
The second monomer, $A_{1}$, can be placed in any one of the $z$ available nearest
neighbour (NN) sites of ${\bf r}_{0}$, chosen at random. Let the walk be non-reversing
so that we have a maximum of $z-1$ sites to choose from for making any further step. Let
$\{ {\bf r}_{k}^{m} \mid m = 1,2, ..., z_{k} \} $ be the 'unoccupied' NN sites available for 
making the $k$th step of the walk. If the number of available sites, $z_{k}$, is zero, then
the walk can not grow further because it is geometrically 'trapped'. In this case, we clear
the lattice and start a fresh walk from ${\bf r}_{0}$ again. If $z_{k} \not= 0$, the walk
proceeds by choosing one of the $z_{k}$ available sites at random with a probability
defined as follows.  

Let $n_{k}^{m}$ be the number of non-bonded NN contacts which the $k$th monomer
of type $A_{k}$ would make with the walk if it were placed at site ${\bf r}_{k}^{m}$. Clearly,
$0 \leq n_{k}^{m}< z-1$. Some of these contacts will be with the H type, and some of
them will be the P type monomers. Let these numbers be denoted by $n_{A_{k}H}^{m}$
and $n_{A_{k}P}^{m}$ respectively. If the energies associated with an HH-contact, an
HP-contact and a PP-contact are denoted by $\varepsilon _{HH}$, $\varepsilon _{HP}$
and $\varepsilon _{PP}$ respectively, then the cost of energy involved in placing the $k$th
monomer of type $A_{k}$ at ${\bf r}_{k}^{m}$ is given by $E_{A_{k}}^{m} =
n_{A_{k}H}^{m}\varepsilon _{A_{k}H}+ n_{A_{k}P}^{m}\varepsilon _{A_{k}P}$.
The probability of choosing the site ${\bf r}_{k}^{m}$ for the $k$th monomer of type $A_{k}$
may now be defined as, $P_{A_{k}}^{m} \equiv $ exp($-\beta E_{A_{k}}^{m}$)/
$\sum _{m}$exp($-\beta E_{A_{k}}^{m}$), where the summation is over all the $z_{k}$
available sites, $\beta = 1/ k_{B}T$, $k_{B}$ is the Boltzmann constant and $T$ the
temperature.

At 'infinite' temperature ($\beta = 0$), the walk is insensitive to the type of monomers being
added and so is identical to the Kinetic Growth Walk (KGW) which in turn is in the same 
universality class as self-avoiding walk [5]. However, at zero temperature ($\beta = \infty $), 
the walk will grow into a compact or an extended conformation depending on whether the 
local site-energies, $E$, are negative or positive respectively. How compact a minimum 
energy conformation is going to be depends strongly on the fraction, $\chi $, of H type 
monomers in the walk - in the limit $\chi \rightarrow 1$, it is the most compact one, whereas 
in the other limit $\chi \rightarrow 0$, it is identical to the KGW which is noncompact. In order 
to mimic the tendency of the H type monomers to minimize contact with the aqueous medium 
by forming clusters, it is customary to choose the contact energies, 
$ \vec{\varepsilon} \equiv (\varepsilon _{HH},\varepsilon _{HP},\varepsilon _{PP}) = (-1,0,0)$, 
in lattice protein models.

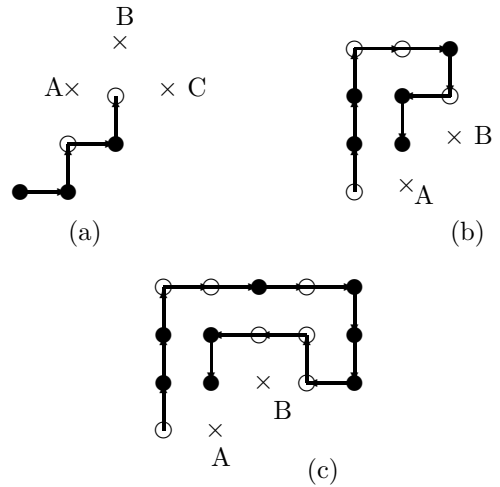
\begin{figure}
\setlength{\unitlength}{0.25in}
\begin{picture}(13,11)
\linethickness{0.075mm}
\linethickness{0.35 mm}
\put(2,7){\circle*{0.35}}

\put(2,7){\vector(1,0){1}}
\put(3,7){\circle*{0.35}}

\put(3,7){\vector(0,1){}}
\put(3,8){\circle{0.35}}

\put(3,8){\vector(1,0){1}} 
\put(4,8){\circle*{0.35}}

\put(4,8){\vector(0,1){}} 
\put(4,9){\circle{0.35}}

\put(2.875,9){$\times$}
\put(2.5,9){A}

\put(3.875,10){$\times$}
\put(4,10.5){B}

\put(4.875,9){$\times$}
\put(5.5,9){C}

\put(3,6){(a)}

\put(9,7){\circle{0.35}}

\put(9,7){\vector(0,1){1}}
\put(9,8){\circle*{0.35}}

\put(9,8){\vector(0,1){1}}
\put(9,9){\circle*{0.35}}

\put(9,9){\vector(0,1){1}} 
\put(9,10){\circle{0.35}}

\put(9,10){\vector(1,0){1}} 
\put(10,10){\circle{0.35}}

\put(10,10){\vector(1,0){1}}
\put(11,10){\circle*{0.35}}

\put(11,10){\vector(0,-1){1}}
\put(11,9){\circle{0.35}}

\put(11,9){\vector(-1,0){1}} 
\put(10,9){\circle*{0.35}}

\put(10,9){\vector(0,-1){1}} 
\put(10,8){\circle*{0.35}}

\put(9.875,7){$\times$}
\put(10.25,6.75){A}

\put(10.875,8){$\times$}
\put(11.5,8){B}

\put(11,6){(b)}

\put(5,2){\circle{0.35}}

\put(5,2){\vector(0,1){1}}
\put(5,3){\circle*{0.35}}

\put(5,3){\vector(0,1){1}}
\put(5,4){\circle*{0.35}}

\put(5,4){\vector(0,1){1}} 
\put(5,5){\circle{0.35}}

\put(5,5){\vector(1,0){1}} 
\put(6,5){\circle{0.35}}

\put(6,5){\vector(1,0){1}}
\put(7,5){\circle*{0.35}}

\put(7,5){\vector(1,0){1}}
\put(8,5){\circle{0.35}}

\put(8,5){\vector(1,0){1}} 
\put(9,5){\circle*{0.35}}

\put(9,5){\vector(0,-1){1}} 
\put(9,4){\circle*{0.35}}

\put(9,4){\vector(0,-1){1}}
\put(9,3){\circle*{0.35}}

\put(9,3){\vector(-1,0){1}}
\put(8,3){\circle{0.35}}

\put(8,3){\vector(0,1){1}} 
\put(8,4){\circle{0.35}}

\put(8,4){\vector(-1,0){1}} 
\put(7,4){\circle{0.35}}

\put(7,4){\vector(-1,0){1}}
\put(6,4){\circle*{0.35}}

\put(6,4){\vector(0,-1){1}}
\put(6,3){\circle*{0.35}}

\put(5.875,1.875){$\times$}
\put(6,1.25){A}

\put(6.875,2.875){$\times$}
\put(7.3,2.25){B}

\put(8,1){(c)}
\end{picture}
\caption{Schematic illustration of HP-chain growth on a square lattice at $T=0$. Hydrophobic and Polar 
monomers are represented by open and filled circles respectively. Suppose the next monomer to be added 
is of Hydrophobic type. (a) Site A will be chosen with probability 1 since it has a NN contact; (b) Sites A 
and B will be chosen at random with probability 1/2 since they both have a NN contact each; (c) Site B
will be chosen with probability 1 since it has two NN contacts whereas site A has only one.}
\label{walk}
\end{figure}

In this case, at $T=0$, an H type monomer will be placed at a site with maximum
HH-contacts whereas a P type monomer is insensitive to the type of NN contacts. We
have schematically illustrated this in Fig. \ref{walk}. It is clear that the total energy associated with
a conformation is equal to the total number of HH-contacts in the walk. However, such
a straightforward adaptation of the IGW algorithm, hereafter referred to as the HP-IGW
algorithm, does not automatically lead to ground state conformations corresponding to a
given HP-sequence. Hence, we have further modified the HP-IGW algorithm so that it
can be used in the multi-pass mode. 

The given HP-sequence, ${\cal S}_{N}$, may be partitioned into $M$ subsequences,
${\cal S}_{N_{1}}, {\cal S}_{N_{2}}, ..., {\cal S}_{N_{M}}$, without changing the
order in which the letters appear in the original sequence. Clearly, ${\cal S}_{N} \equiv 
{\cal S}_{N_{1}}\cup {\cal S}_{N_{2}}\cup ...\cup {\cal S}_{N_{M}}$ with $N = 
N_{1}+N_{2}+ ...+N_{M}$. We concentrate on the zero temperature case.

In 'pass' 1, we use the HP-IGW algorithm to generate a large number of chains consisting
of $N_{1}$ monomers in the order, as well as of the type, specified by the first subsequence, 
${\cal S}_{N_{1}}$. Out of all the conformations generated, we store only those with
minimum energy, say $-{\cal E}_{1}$. Let ${\cal N}_{1}$ denote the ensemble as well as 
the number of all these minimum energy conformations. This part of the algorithm is not
truely kinetic because it does not simulate the relaxation kinetics in the conformation space.
Nevertheless, together with the chain growth algorithm, it quickly leads to the ground
state conformations. 

In 'pass' 2, we take a chain, say $C_{1}^{1} \in {\cal N}_{1}$, and use the HP-IGW 
algorithm to let it grow further by a chain segment consisting of $N_{2}$ monomers whose 
order and type are specified by the next subsequence, ${\cal S}_{N_{2}}$. Thus we obtain a chain 
consisting of $N_{1}+N_{2}$ monomers whose type and order are specified by the subsequence, 
${\cal S}_{N_{1}} \cup {\cal S}_{N_{2}}$. We use the same chain segment $C_{1}$ again 
and again  for generating a specified number of $(N_{1}+N_{2})$-monomer chain segments.
We repeat this for all the chain segments, $\{ C_{j}^{1}\in {\cal S}_{N_{1}}\mid j = 1,2, ..., N_{1}\}$.
It is clear that all these chain conformations will have energy less than or equal to $-{\cal E}_{1}$.
Let ${\cal N}_{2}$ denote the ensemble as well as the number of $(N_{1}+N_{2})$-monomer 
conformatins with minimum energy, say $-{\cal E}_{2}$. 

\begin{figure}
\setlength{\unitlength}{0.5mm}
\begin{picture}(130,110)
\linethickness{0.075mm}

\multiput(30,15)(30,0){3}{\line(0,1){100}}
\put(20,5){Pass 1}
\put(50,5){Pass 2}
\put(80,5){Pass 3}

\put(0,110){\line(1,0) {30}}
\put(0,110){\line(6,-1) {30}}
\put(0,110){\line(3,-1) {30}}
\put(0,110){\line(2,-1) {30}}

\put(30,95){\line(1,0) {30}}
\put(30,95){\line(6,-1) {30}}
\put(30,95){\line(3,-1) {30}}
\put(30,95){\line(2,-1) {30}}

\put(60,80){\line(1,0) {30}}
\put(60,80){\line(6,-1) {30}}
\put(60,80){\line(3,-1) {30}}
\put(60,80){\line(2,-1) {30}}

\put(120,100){\vector(0,-1){50}}
\put(120,102){$0$}
\put(115,45){$-{\cal E}$}

\put(15,90){$-{\cal E}_{1}$}
\put(45,75){$-{\cal E}_{2}$}
\put(75,60){$-{\cal E}_{3}$}

\put(32,98){${\cal N}_{1}$}
\put(62,83){${\cal N}_{2}$}
\put(92,68){${\cal N}_{3}$}
\end{picture}
\caption{Schematic illustration of the muti-pass algorithm described in the text. ${\cal N}_{1}$,
${\cal N}_{2}$ and ${\cal N}_{3}$ are the ensembles of conformations with minimum energies,
$-{\cal E}_{1}$, $-{\cal E}_{2}$ and $-{\cal E}_{3}$ at the end of passes 1,2 and 3 respectively.}
\label{multipass}
\end{figure}
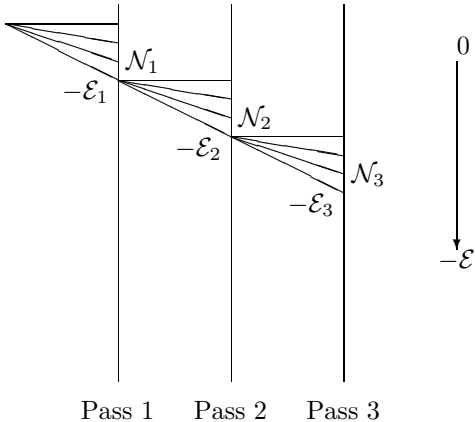

Similarly, in 'pass' 3, we generate an ensemble, ${\cal N}_{3}$, of $(N_{1}+N_{2}+N_{3})$-monomer
conformations with minimum energy $-{\cal E}_{3}$, and so on.  Thus, at the end of 'pass' M, we 
have an ensemble of $N$-monomer chain conformations with minimum energy, 
$-{\cal E} (<-{\cal E}_{M-1}< ...<-{\cal E}_{2}<-{\cal E}_{1})$. It may be noted that the order and 
type of the monomers in the chains generated are as specified by the original sequence, 
${\cal S}_{N}$. We have schematically illustrated this algorithm in Fig.\ref{multipass}.
We generate a large number of chain conformations at every stage, not for the purpose of statistics, but to 
make sure that we obtain as many distinct conformations as possible. This improves the chances 
of obtaining minimum energy states. Moreover, the optimal lengths of the subsequences ({\it i.e.,}
the optimal values of $N_{1}, N_{2}, N_{3}$ {\it etc.,}) for which we may get the 'global' minimum 
are strongly dependent on the particular sequence under consideration.

We have tested this multi-pass algorithm for all the ten bench mark sequences of Yue {\it et al} [6]
and for two of Toma $\&$ Toma [7]. These sequences were designed by minimizing the energy of a 
particular target conformation on a cubic lattice with the energy parameters given by 
$\vec{\varepsilon}=(-1,0,0)$. We spent a maximum of thirty minutes CPU time on a DEC Alpha 
Workstation for each of these bench mark sequences.We could obtain the ground state conformations 
for some of these sequences in just two passes - namely, the first halves of the chains in the first pass, 
and the second halves in the second pass - while for others, in three passes. We find that generating 
the first segment of the chain from an arbitrary point in the sequence, rather than always from the 
beginning, ultimately leads to better ground states:  ({\it i}) we occupy the 
site ${\bf r}_{0}$ with the monomer of the type specified by the $n (1 \leq n \leq N_{1})$th letter in the 
subsequence ${\cal S}_{N_{1}}$; ({\it ii}) the next monomer will then correspond either to the 
$(n-1)$th or to the $(n+1)$th letter in the sequence with probability 1/2, and so on. 
In other words, the chain will grow either to the left or to the right at random 
with equal probability. However, the successfully grown chain will finally have monomers in the same 
order as well as of the type specified by the subsequence, ${\cal S}_{N_{1}}$. The chain 
segments in subsequent passes will be grown always in the order in which they should grow. We handled 
these passes separately and manually, rather than as integrated modules in a single program.
In Table \ref{energy}, we have presented our results along with the ones reported in the literature.

\begin{table}[t]
\caption{Ground state energies obtained for the $3d$ bench mark sequences. All the sequences of
Yue {\it et al.} [6] are of length $N=48$. The sequences 10 and 11 of Toma and Toma [7] are of lengths,
$N=46$ and $N= $ respectively. The two energy values presented in the middle column are those of
Yue {\it et al} and Bastolla {\it et al} [6]}
\label{energy}
\vspace{0.1in}
\begin{tabular}{clcc} 
Ref. & Seq. & $-E_{min}$(reported) & $-E_{min}(ours)$\\ \hline\hline  
&&&\\
Yue {\it et al} [6]        &   1            & 31, 32 & 31\\
                                  &   2            & 32, 34 & 32\\
                                  &   3            & 31, 34 & 32\\
                                  &   4            & 30, 33 & 30\\
                                  &   5     	& 30, 32 & 30\\
                                  &   6            & 30, 32 & 30\\
                                  &   7            & 31, 32 & 31\\
                                  &   8 	& 31, 31 & 30\\
                                  &   9            & 31, 34 & 31\\
                                  & 10            & 33, 33 & 31\\ \hline 
&&&\\
Toma $\&$ Toma [7] & 10	& 34 & 33\\
                                  & 11            & 42 & 41\\ 
\end{tabular}
\end{table}

In Fig.3, we have shown the distributions of NN contacts corresponding to the three passes used for Sequence 3 
of Yue {\it et al}. Configurations of the first segment of length, $N_{1} = 19$, with maximum number of
contacts ($n_{max}=9$) were used in the second pass for growing the second segment of length, $N_{2} = 17$, 
whose minimum energy configurations ($n_{max}=22$)
\begin{figure}
\includegraphics[width=3.25in,height=2.4in]{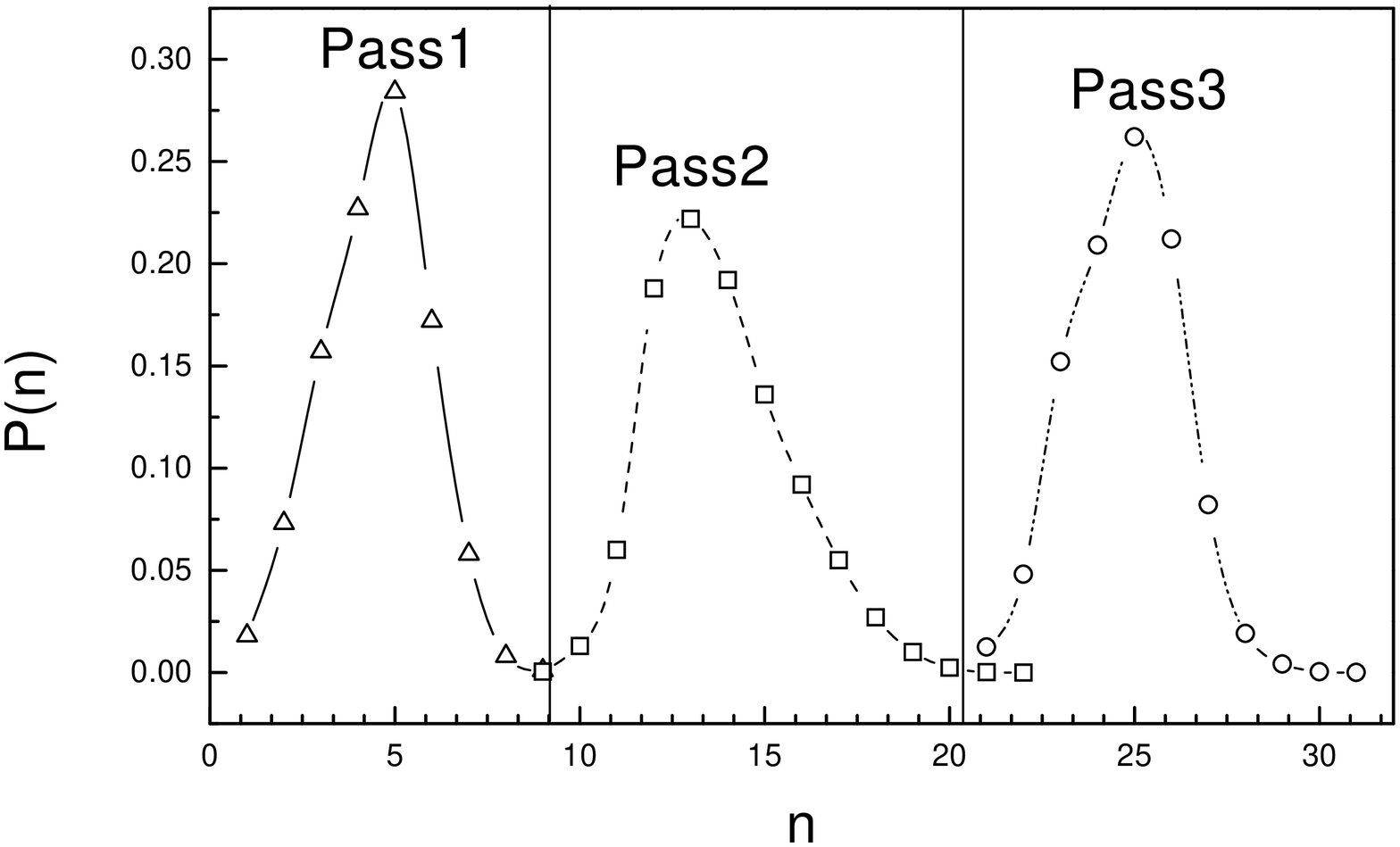}
\caption{The distributions of NN contacts obtained in the three passes used for the Sequence 3 of Yue {\it et al} } 
\end{figure}
\newpage

\noindent  were in turn used for growing the last segment of length, 
$N_{3} = 12$, in the third pass. The distributions, not normalized in any way with respect to each other, 
schematically illustrate how the
number of contacts in a typical configuration can be increased in a progressive manner. 

Thus, we have a simple but powerful growth algorithm which can be used for obtaining the ground state 
conformations of lattice proteins. A finite temperature version of this algorithm can be implemented in two 
ways. Since the temperature used in the chain growth process, say $T_{G}$, could be different from the 
temperature $T$ at which the fully grown conformations are sampled, we may use the growth module 
HP-IGW either at $T_{G}=0$ or at $T_{G} \not= 0$. In the latter case, we may set $T_{G}=T$ 
so as to have just one temperature for the entire process. Having generated a chain conformation  
of energy ${\cal E}$, we may store it with a probability proportional to the Botzmann factor, 
exp($-\beta{\cal E})$. This algorithm is similar to the simple enrichment scheme of  Wall and Erpenbeck [8] 
for KGW at $T_{G}=T \rightarrow \infty$. It is also likely that this finite temperature version is more efficient 
for a ground state search than the zero temperature version discussed in this paper. A detailed study of this 
algorithm and its variants will be reported elsewhere.

S. L. N thanks A. K. Rajarajan for critical reading of the manuscript.

\noindent $^*$slnoo@magnum.barc.ernet.in;


\begin{references}
\bibitem{1} C. B. Anfinsen, Science, {\bf 181}, 223 (1973); J. N. Onuchic, Z. A. Luthey-Schulten
and P. G. Wolynes, Ann. Rev. Phys. Chem. {\bf 48}, 545 (1997).
\bibitem{2} D. Baker, Nature, {\bf 405}, 39-42 (2000).
\bibitem{3} K. A. Dill, Biochemistry, {\bf 24}, 1501 (1985); H. Taketomi, Y. Ueda and N. Go,
Int. J. Pept. Protein Res. {\bf 7}, 445 (1975); K. F. Lau and K. A. Dill, Macromolecules, {\bf 22},
3986 (1989); J. Chem. Phys. {\bf 95}, 3775 (1991); D. Shortle, H. S. Chan and K. A. Dill,
Protein Sci. {\bf 1}, 201 (1992); J. D. Bryngelson {\it et al}, Proteins, {\bf 21}, 167 (1995);
K. A. Dill, S. Bromberg, K. Yue, K. M. Feibig, D. P. Yee, P. D. Thomas and H. S. Chan, 
Protein Sci. {\bf 4}, 561 (1995); D. K. Klimov and D. Thirumalai, Phys. Rev. Lett. {\bf 76},
4070 (1996); E. Shakhnovich, Curr. Opin. Struct. Biol. {\bf 7}, 29 (1997); V. S. Pande {\it et al},
Curr. Opin. Struct. Biol. {\bf 8}, 68 (1998).
\bibitem{4} H. Frauenkron, U. Bastolla, E. Gestner, P. Grassberger and W. Nadler, Phys. Rev. Lett.
{\bf 80}, 3149 (1998); G. Chickenji, M. Kikuchi and Y. Iba, Phys. Rev. Lett.
{\bf 83}, 1886 (1999); R. Unger and J. Moult, J. Mol. Biol. {\bf 231}, 75 (1993); U. H. E. Hansmann
and Y. Okamoto, J. Comp. Chem. {\bf 14} 1333 (1993); Physica, {\bf A212}, 415 (1994);
Phys. Rev {\bf E54}, 5863 (1996); A. Sali and E. I. Shaknovich, J. Mol. Biol. {\bf 235}, 1614 (1994).
\bibitem{5} S. L. Narasimhan, P. S. R. Krishna, K. P. N. Murthy and M. Ramanadham, Phys. Rev. E,
Rapid Comm. in press (2001).
\bibitem{6} K. Yue {\it et al}, Proc. Natl. Acad. Sci. U. S. A, {\bf 92}, 325 (1995); U. Bestolla, 
H. Frauenkron, E. Gerstner, P. Grassberger and W. Nadler, Proteins: Struct. Func. Genetics, {\bf 32}, 52 (1998);
\bibitem{7} L. Toma and S. Toma, Protein Sci. {\bf 5}, 147 (1996).
\bibitem{8} F. T. Wall and J.J. Erpenbeck, J. Chem. Phys. {\bf 30}, 634 (1959); {\it ibid.}, 637 (1959). 

\end{references}
\end{document}